# Lateral distribution of Cherenkov light in extensive air showers at high mountain altitude produced by different primary particles in wide energy range

Alexander Mishev[*] and Jordan Stamenov

*Institute for Nuclear Research and Nuclear Energy, Bulgarian Academy of Sciences, 72, Tsarigradsko chaussee, Sofia 1284, Bulgaria*

**Abstract:** The general aim of this work is to obtain the lateral distribution of atmospheric Cherenkov light in extensive air showers produced by different primary particles precisely by. Protons, Helium, Iron, Oxygen, Carbon, Nitrogen, Calcium, Silicon and gamma quanta in wide energy range at high mountain observation level of 536 g/cm$^2$ - Chacaltaya cosmic ray station. The simulations are divided generally in two energy ranges $10^{10}$ eV–$10^{13}$ eV which can be used for ground based gamma ray astronomy and $10^{13}$ - $10^{17}$ eV which can be used for wave front sampling telescope arrays experiments. One large detector with dimensions 800x800m has been used for simulations, the aim being to reduce the statistical fluctuations of the obtained characteristics. The shape of the obtained lateral distributions of Cherenkov light in extensive air showers is discussed and the scientific potential for solution of different problems as well.

## 1. Introduction

This field of astroparticle physics is connected with gamma-ray astrophysics and they both are young and dynamic. This field crossroads the particle physics and astrophysics, is rapidly extending in the few last years. In the last several decades one has been able to observe phenomena such as supernova remnants, active galactic nuclei, gamma ray bursts which have significant impact to our knowledge of the universe. At the same time there are still several very important unsolved problems connected with the origin and acceleration mechanisms of primary cosmic ray flux. In fact the cosmic ray studies are complementary to gamma ray astrophysics since many gamma rays are produced in processes connected with cosmic ray such as synchrotron emission as example, which involve charged cosmic ray particles.

The measurements of the individual cosmic ray spectrum and the precise estimation of mass composition are very important in attempt to obtain more detailed information about the sources of primary cosmic ray and to build an adequate model of cosmic ray origin [1].

Above $10^{14}$ eV the possibility for cosmic ray detection and measurement is ground based i.e. the detection of the secondary cosmic ray as example the Cherenkov light in extensive air shower (EAS). One of the most convenient techniques in cosmic ray investigation is the atmospheric Cherenkov technique [2]. At the same time it is important to cover the gap between the ground based and the space-born gamma ray astronomy. Currently gamma-ray energies between 20 and 250 GeV are not accessible to space-borne detectors and ground-based air Cherenkov detectors, the exception is the new generation imaging air-Cherenkov telescopes such as HESS [3, 4]. The air Cherenkov telescopes based on the image technique have become the most power-full instrument for ground based gamma ray astronomy. The typical threshold for the majority of the telescopes ranges around 1 TeV. After the discovery of many new gamma ray sources by EGRET at COMPTON observatory the construction of new ground based telescopes with aim to decrease the energy threshold was started. Several in preparation experiments such as TACTIC [5] or MAGIC [6] used the image technique i.e. the reconstruction of the Cherenkov image of the shower. The scientific potential of the ground based gamma ray astronomy is enormous and covers both astrophysics and fundamental physics as was mentioned above. In one hand it is possible to study objects such as supernova remnants, active galactic nuclei and pulsars. On the other hand the observations especially in the range of low energies will help to understand well the various acceleration mechanisms assumed to be at the origin of very high energy gamma quanta.

---

[*] Corresponding author: mishev@inrne.bas.bg

The registration of the atmospheric Cherenkov light in EAS can be applied for both of the cited above problems i.e. the study of the mass composition and energy spectrum of the primary cosmic ray and the gamma astronomy.

The detection of the air Cherenkov light at ground level using an array of telescopes or photomultipliers contrary to the image technique is also a powerful tool for the both of the mentioned above problems - gamma astronomy [7, 8] and the all particle energy spectrum [9, 10].

At the same time the new telescopes in development needs an accurate and detailed analysis of their performances i.e. the calculation of the detector response. This is possible on the basis of Monte Carlo simulations. Among the several codes on the market the CORSIKA code [11] has become practically the standard in cosmic ray community. In fact the most of the published data so far includes the detector features and constraints for which the simulations are carried out and therefore in the case when one needs the lateral distribution of Cherenkov light in EAS for different experiment it is obvious the need to carry out the simulation once more separately. Additionally the corresponding observation levels are different for the different experiments in preparation. With this in mind we present several systematic results based on Monte Carlo simulations with help of CORSIKA 6.003 code [11]. Moreover the presented results i.e. the lateral distribution of Cherenkov light in EAS produced by different primaries at Chacaltaya observation level of 536g/cm$^2$ have been used for basis on a previously proposed method for primary cosmic ray mass composition estimation and energy spectrum estimation [12, 13, 14] and ground based gamma ray astronomy [14, 15, 16] developed for HECRE experimental proposal [17].

Generally the results can be interpreted for different problems solution in two energy ranges. The first energy range is between $10^{10}$ eV and $10^{13}$ eV can be used for ground based gamma astronomy i.e. for separation of gamma initiated showers from hadron initiated events. Such type of experiments are the wave front sampling telescopes with large mirrors or solar power plants (a good example is STACEE [18, 19]). The second energy range between $10^{10}$ eV and $10^{17}$ eV can be used for wave front and angle integrating telescope arrays and as was discussed above can be used for all particle spectrum estimation of primary cosmic ray around the "knee". It is clear that the obtained lateral distributions of Cherenkov light in EAS can be used for check of the proposed reconstruction techniques based on different methods. This will permit to estimate the different constraints and advantages of the previously proposed method [12, 14] and to compare with different methods. The results presented in this work are useful for experiments in preparation taking into account that the detector response is not included in present simulations nor the constraints connected with the detector electronics.

One of the main problems during the calculations was the huge computational time necessary for tracking the charged particles which are above the energy threshold of the Cherenkov effect in the atmosphere. This is one of the reasons to choose the high mountain observation level for the simulations in attempt to reduce in one hand the computational time and on the other hand to reduce the fluctuations of the obtained distributions because this level is near to the showers maximum. At the same time at this observation level is planed the development of new experiment according the HECRE proposal [17].

In addition the presented results of the simulation are a good basis for checking the previously proposed selection parameter [20, 21] used for constant efficiency selection of the registered events for other primaries different from proton and gamma quanta.

## 2. The Simulation

The CORSIKA code [11] version 6.003 has been used for simulation of the development of EAS precisely the atmospheric Cherenkov light. The GHEISHA [22] and QGSJET [23] hadronic interaction models have been used respectively for low and high energy hadronic interactions. The observation level was of 536g/cm$^2$ which is near to the shower maximum. As a result the fluctuation in the shower development are not so important comparing to lower observation levels and it is possible to obtain flatter distributions of the different shower components precisely the lateral distribution of Cherenkov light in EAS. The simulated particles are primary protons, Iron, Helium, Oxygen, Carbon, Nitrogen, Silicon and Calcium nuclei and primary gamma quanta as well. The lateral distribution of the Cherenkov light flux in EAS was obtained using one large detector of 800x800m the aim to reduce the statistical fluctuations and to collect as much as possible of the Cherenkov photons in the shower. The detector is divided in 23 bins distributed in logarithmic scale. This is important for the further approximation of the obtained lateral distribution taking into account several advantages of the used method [24, 25] and the REGN code [26].

The most important fluctuations in shower development are due to the longitudinal shower development. In one hand the tracking of the totality of the Cherenkov photons individually in the shower is quite difficult because the enormous amount of computational time [27, 28]. On the other hand using one large detector to collect practically the totality of Cherenkov photons at given observation level is not possible because the enormous disk space. As example one proton induced shower of $10^{13}$ eV at Chacaltaya observation level takes at lest 2 Gb disk space using for registration the cited above detector. Thus we decide to sample the simulation using the well known procedure existing in the CORSIKA code i.e. the Cherenkov photons are grouped in bunches [29]. Obviously the bunch size depends on the energy of the incident primary particle. We apply the CORSIKA algorithm for automatic bunch size calculation. In one hand the procedure is most computing consuming in comparison to give some value to the bunch. On the other hand this procedure is practically the optimal [11, 30] In the low energy region i.e. for gamma quanta and protons above $10^{10}$ eV the calculated bunch size is 30 photons. In the high energy region the calculated bunch size is 177 photons for $10^{15}$ eV proton induced events, 1834 for $10^{16}$ proton induced events and 18405 for $10^{17}$ eV proton induced events. The bandwidth of Cherenkov photons simulations was chosen between 300nm and 550 nm, which is the sensitivity range of the most of photomultipliers in production and which are used in the different experiments counting on atmospheric Cherenkov technique.

The Rayleigh attenuation of the light in the atmosphere was not taken into account, nor the Mie scattering. One of the reasons to make this is that these effects are easy to calculate and to take into account [31] and at this observation level the amount of the atmosphere above the observation level is not so important. The consequence is the possibility to neglect these effects. Only vertical events are taken into account during the simulations. The used statistics varies as a function of the energy of the incident particle. In low energy region we simulate at least 10 000 events per energy point. Above $10^{13}$ eV we simulate 500 events per energy point. The core of each is located at the center of the detector. Thus we simulate showers of the same type.

In attempt to save disk space and to use large statistics a partial modification of the original CORSIKA version was made, precisely in the code output. All the characteristics of interest are calculated during the simulation. Therefore it is possible to obtain event per event directly the needed Cherenkov photon density. The bins are as follows (0., 1., 1.33, 1.78, 2.37, 3.16, 4.22, 5.62, 7.5, 10., 13.3, 17.8, 23.7, 31.6, 42.2, 56.2, 75.,100., 133., 178., 237., 316., 422., 562. ) in meters from the shower axis.

## 3. Results and Discussion

The obtained with CORSIKA 6.003 [11] code lateral distributions of Cherenkov light in EAS at high mountain altitude using corresponding hadronic interaction models GHEISHA [22] and GGSJET [23] are presented in next several figures. In fig. 1a and 1b are presented the lateral distribution of Cherenkov light in EAS produced by proton incoming showers in the energy ranges $10^{11}$-$10^{13}$ eV and $10^{13}$-$10^{17}$ eV at 536g/cm$^2$ observation level. In Y axis is shown the Cherenkov light density Q(R) measured in photons per m$^2$. In X axis is shown the core distance in meters.

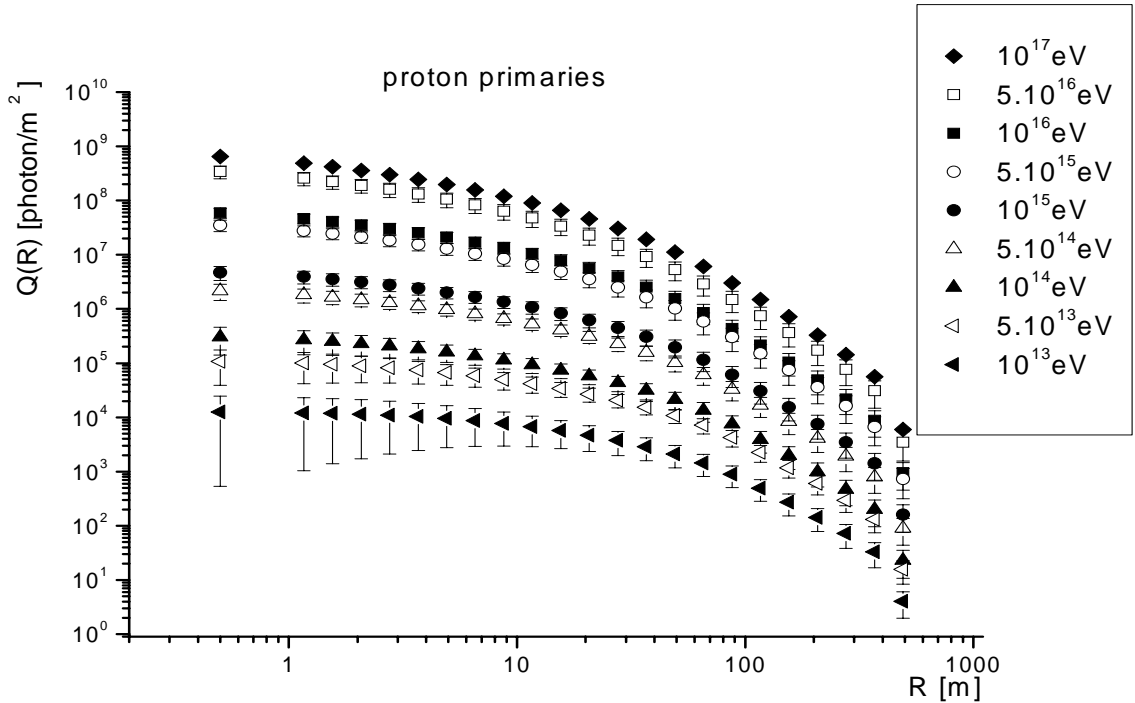

**Fig. 1a** Lateral distributions of Cherenkov light flux in EAS produced by primary protons induced showers in the energy range $10^{13}$ –$10^{17}$ eV at 536g/cm$^2$ observation level

In this figure are presented the mean values with the corresponding standard deviation. An additional analysis in each bin concerning the symmetry of the obtained distribution is carried out. As was expected [29] around $10^{13}$ eV energy of the primary particle and near to the shower axis the distribution of the Cherenkov photons in the bin is not symmetric. As a consequence one observes difference between the mean and median values. This difference is presented in fig. 2 for proton induced showers of $10^{13}$ eV energy. We analyze also each bin per bin. In the energy range $10^{11}$-$10^{13}$ eV the fluctuations distribution of the Cherenkov photons in the bin are higher than from Poisson distribution. Increasing the energy of the incident particle the fluctuations of the Cherenkov photon distribution in the bin became smaller and the distribution is in parctice symmetric (fig.3 and fig.4). The same effect was observed at large distances from the shower axis. This is due to the large bin size at big distance from the shower axis.

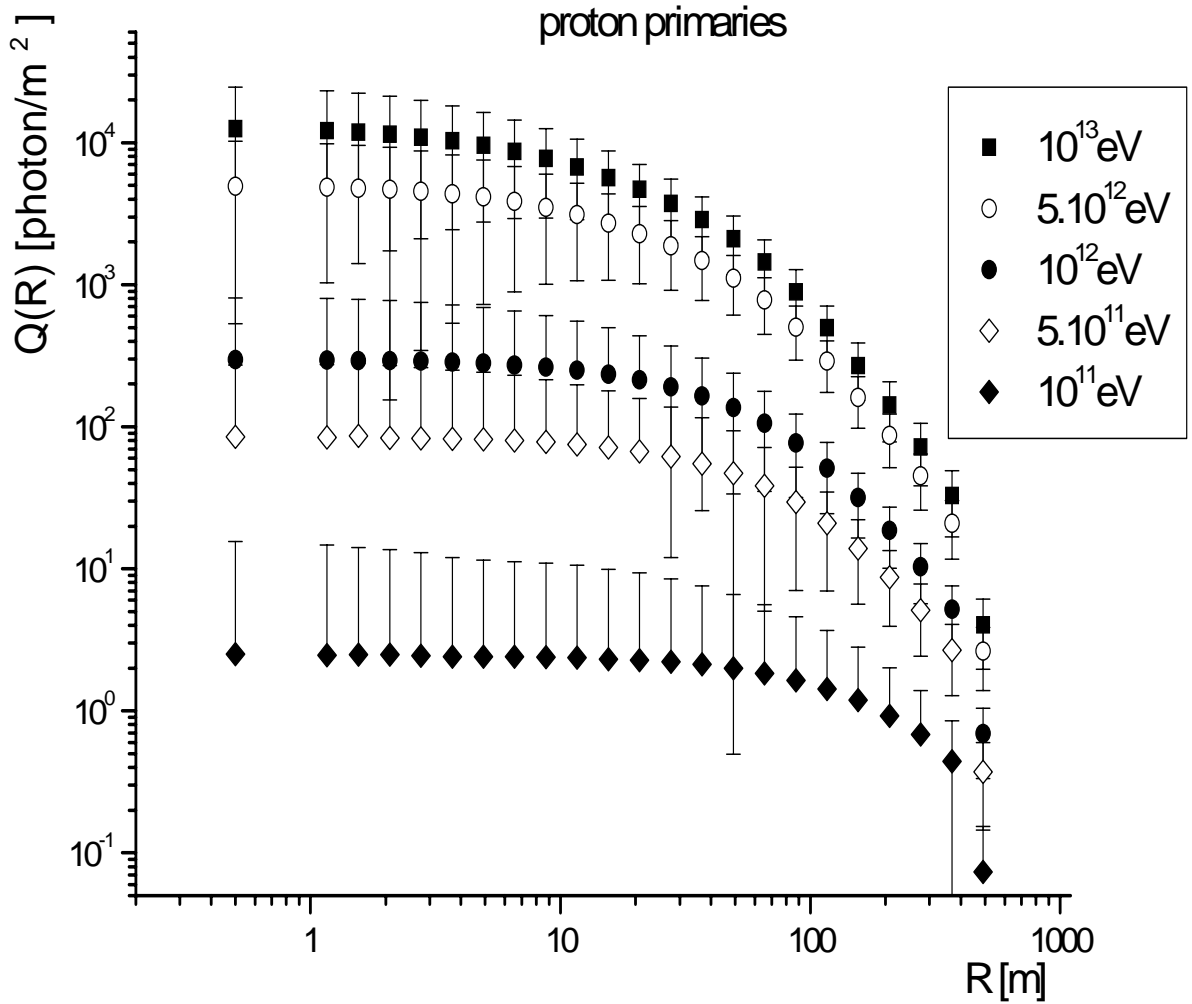

**Fig. 1b** Lateral distributions of Cherenkov light flux in EAS produced by primary protons induced showers in the energy range $10^{11}$ –$10^{13}$ eV at 536g/cm$^2$ observation level

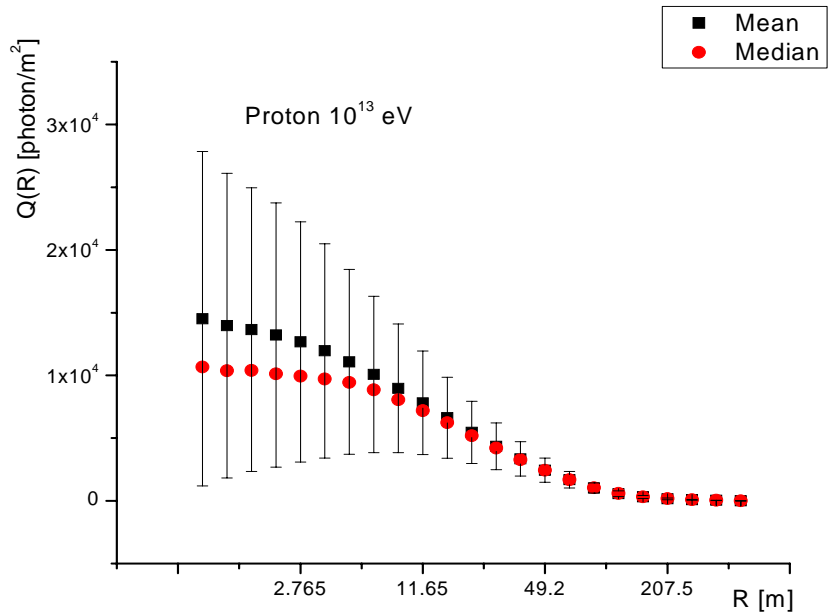

**Fig. 2** The mean and median of the lateral distribution of Cherenkov light flux in EAS produced by primary protons induced showers with energy $10^{13}$ eV at 536g/cm$^2$ observation level

**Fig. 3** The mean and median of the lateral distribution of Cherenkov light flux in EAS produced by primary protons induced showers with energy $10^{15}$ eV at 536g/cm$^2$ observation level

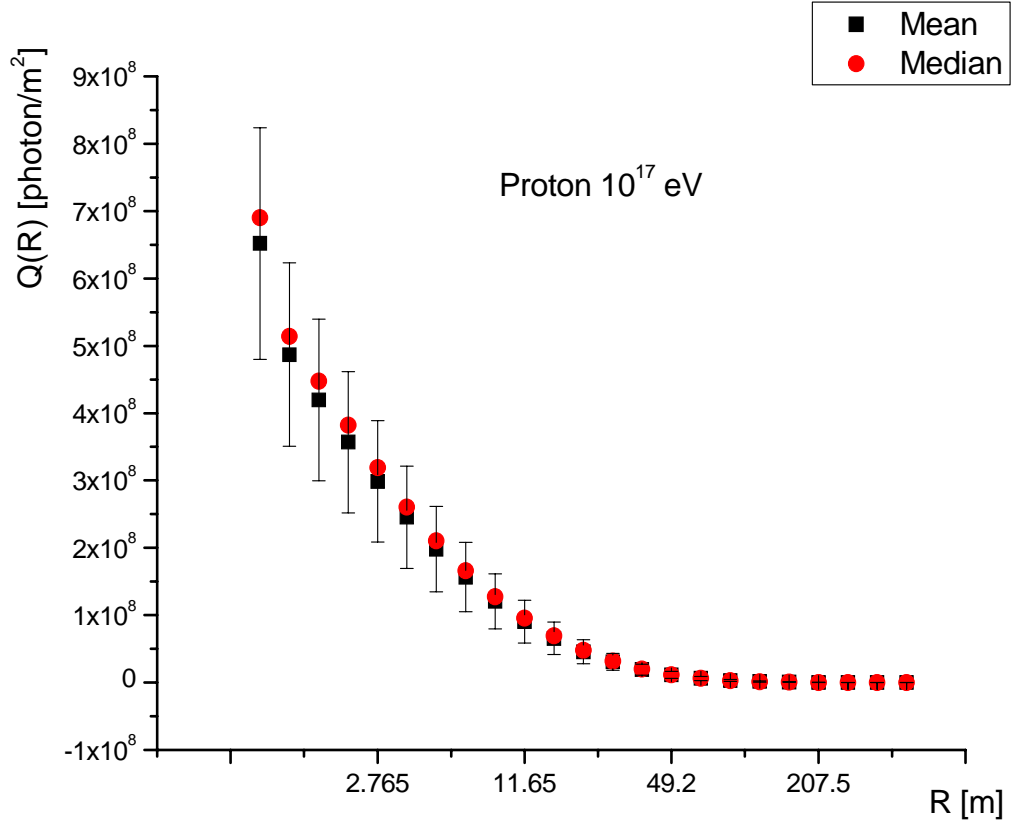

**Fig. 4** The mean and median of the lateral distribution of Cherenkov light flux in EAS produced by primary protons induced showers with energy $10^{17}$ eV at 536g/cm$^2$ observation level

In addition we study the dependence of the fluctuations in Cherenkov light flux as a function of the bin size. This very important especially for low energy region where the obtained densities are lower and the observed fluctuations of the Cherenkov light flux are bigger. The result of this study is that in low energy range the statistical fluctuations are strongly dependent on the size of the bin. This means that in this energy range one have to increase the obtained statistical fluctuations when one deal with detectors having smaller area. As a consequence this result could reflect on the possible reconstruction techniques of the measured event using wave front sampling telescope arrays. It is possible to underestimate the measured Cherenkov light flux in EAS fluctuations and to estimate as example the energy of the reconstructed particle with not correct accuracy. Moreover it is possible to obtain not correct rejection of hadronic induced events when one deal with ground based gamma-ray experiments. Therefore in the low energy region between $10^{10}$ and $10^{13}$ eV one have to study the problems additionally and to use the presented data of simulations only on methodological level.

In the high energy range the fluctuations due to the longitudinal development of the shower dominate and the dependence of the statistical fluctuations is not so strong as a function of the bin size. One can see that at large distances the median of the obtained distribution and the mean values are very close when one increases the bin size and the energy of the incoming primary particle.

This is the reason to present in the figures the mean values of Cherenkov light flux density with the corresponding standard deviation. In fig. 5 is presented the lateral distribution of Cherenkov light in EAS produced by Iron nuclei as incoming showers in the energy range $10^{13}$-$10^{17}$ eV at 536g/cm$^2$ observation level. In fig. 6 is presented the lateral distribution of Cherenkov light in EAS produced by Helium and Oxygen nuclei as incoming showers in the same energy range of $10^{13}$-$10^{17}$ eV at 536g/cm$^2$ observation level.

The additional result is the obtained the influence of the bunch size to the obtained statistical fluctuations of the Cherenkov light density in EAS. Even in the high energy range the intrinsic fluctuations in lateral distribution of Cherenkov light in EAS are grater to the fluctuations due to the changes in a bunch size.

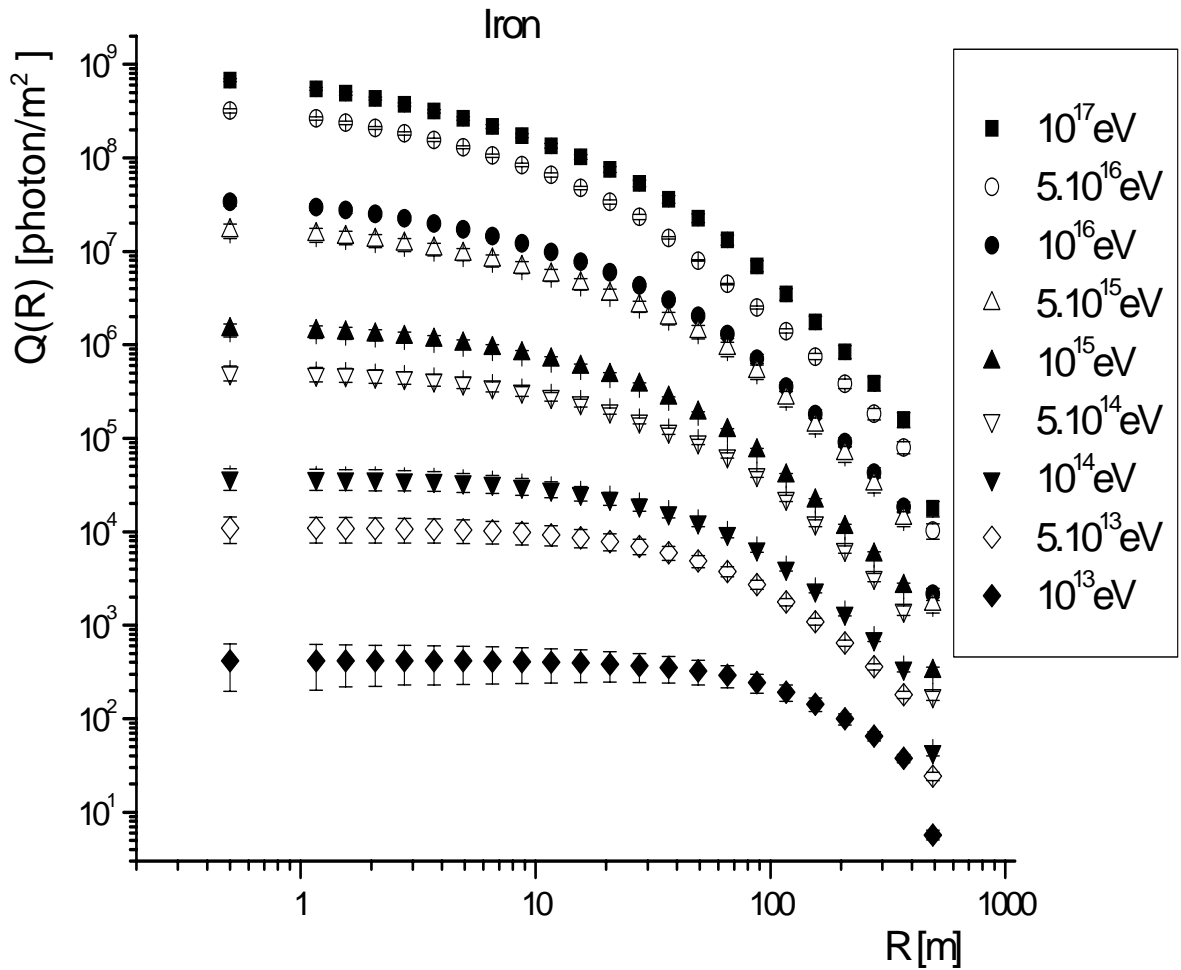

**Fig. 5** Lateral distributions of Cherenkov light flux in EAS produced by primary Iron nuclei induced showers in the energy range $10^{13}$ –$10^{17}$ eV at 536g/cm$^2$ observation level

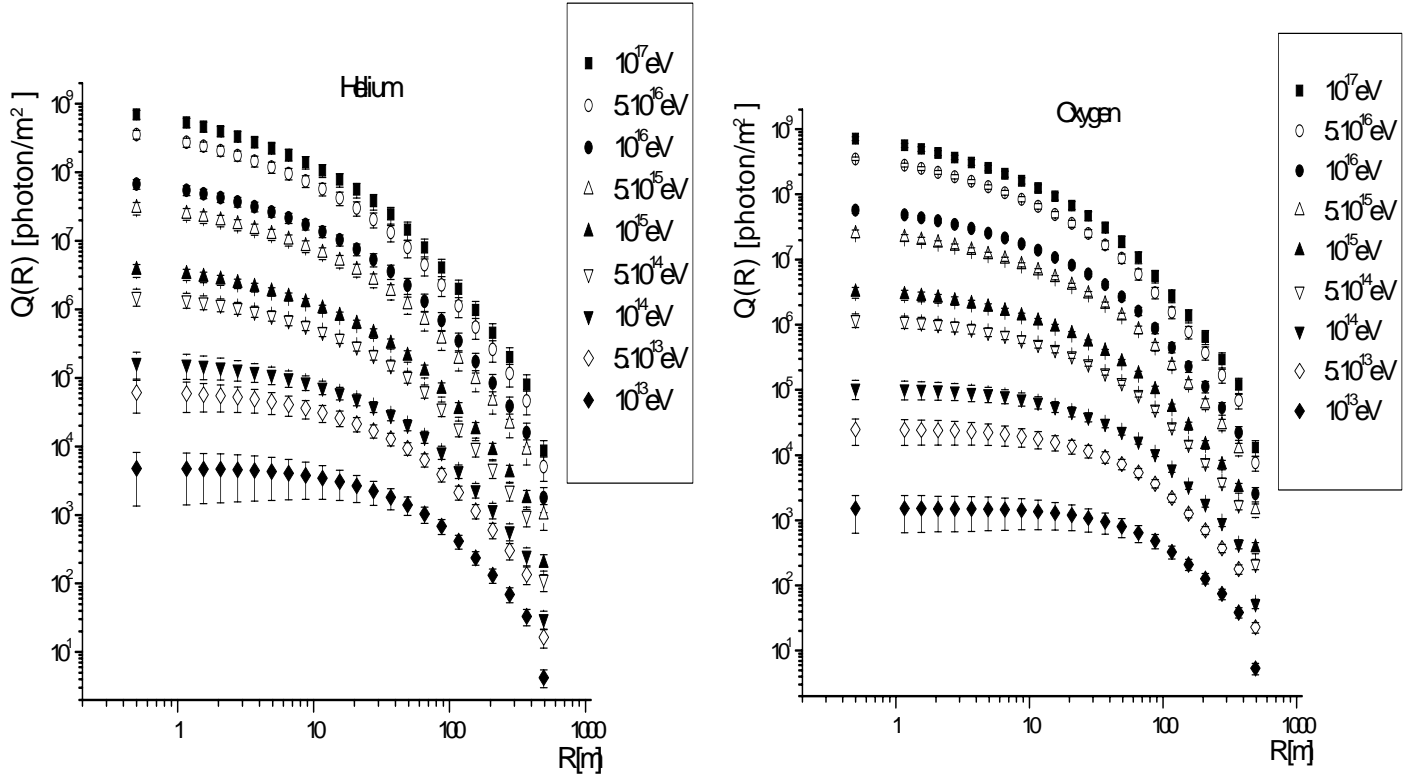

**Fig. 6** Lateral distributions of Cherenkov light flux in EAS produced by primary Helium and Oxygen nuclei induced showers in the energy range $10^{13}$ –$10^{17}$ eV at 536g/cm$^2$ observation level

In all the plots the open symbols was used for lateral distribution of Cherenkov light flux when the energy of the initiated particle is a half order. As example in fig.1 the presented lateral distribution of Cherenkov light in EAS produced by primary proton having energy of $5.10^{15}$ eV is presented with open circles. The filled symbols are used when the energy of the incident particle is equal exactly to the order (as example $10^{15}$ eV is presented with filled circles in the same figure).

It is obvious that the lateral distributions of Cherenkov light initiated in EAS produced by different primary particles are with very similar shape. As was expected the fluctuations are more important at the energies till $10^{13.5}$ eV, thus in low energy range of simulated events. Generally the lateral distribution of Cherenkov light initiated by primary nuclei is wider. Moreover increasing the atomic mass of the incoming particle the lateral distribution became wider and as was expected increasing the atomic mass A of the incoming particle the resulting Cherenkov light flux in EAS decreases. Additionally the fluctuations of the obtained lateral distribution of Cherenkov light in EAS diminish increasing the atomic mass A of the initiating primary particle. In fig. 7 are presented the lateral distributions of Cherenkov light flux in EAS produced by primary Protons and Iron, Helium and Oxygen nuclei induced showers in the energy range $10^{15}$ – $10^{17}$ eV at 536g/cm$^2$ observation level near to the shower axis.

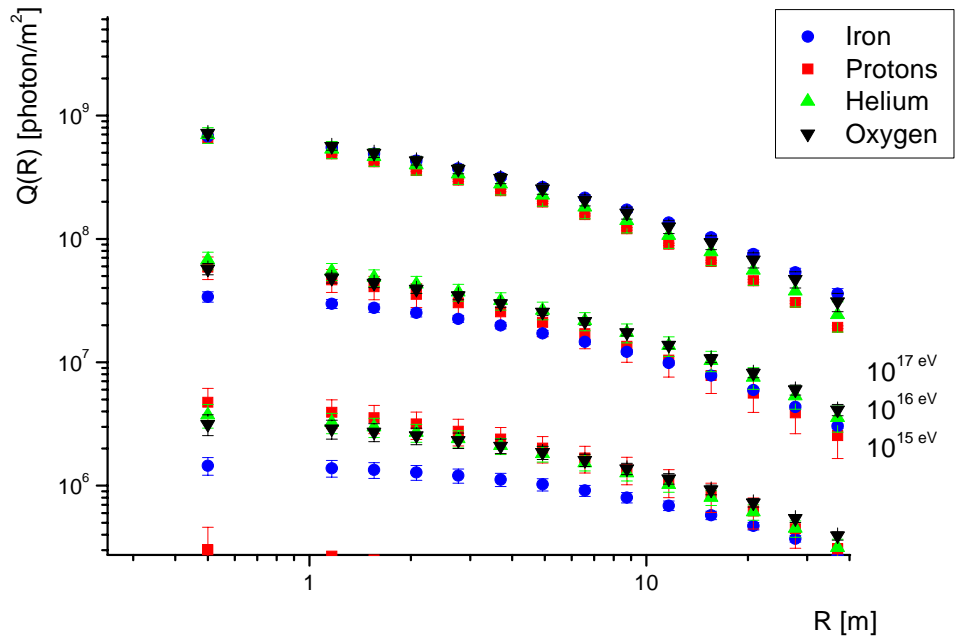

**Fig. 7** Lateral distributions of Cherenkov light flux in EAS near to the shower axis produced by primary Protons and Iron, Helium and Oxygen nuclei induced showers in the energy range $10^{15}$ – $10^{17}$ eV at 536g/cm$^2$ observation level

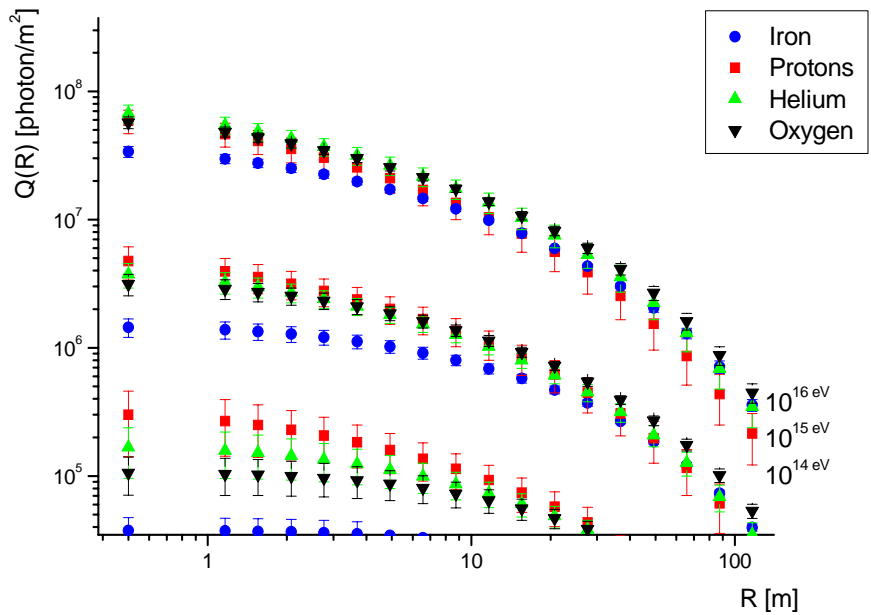

**Fig. 8a** Lateral distributions of Cherenkov light flux in EAS produced by primary Protons and Iron, Helium and Oxygen nuclei induced showers in the energy range $10^{15}$ –$10^{17}$ eV at 536g/cm$^2$ observation level

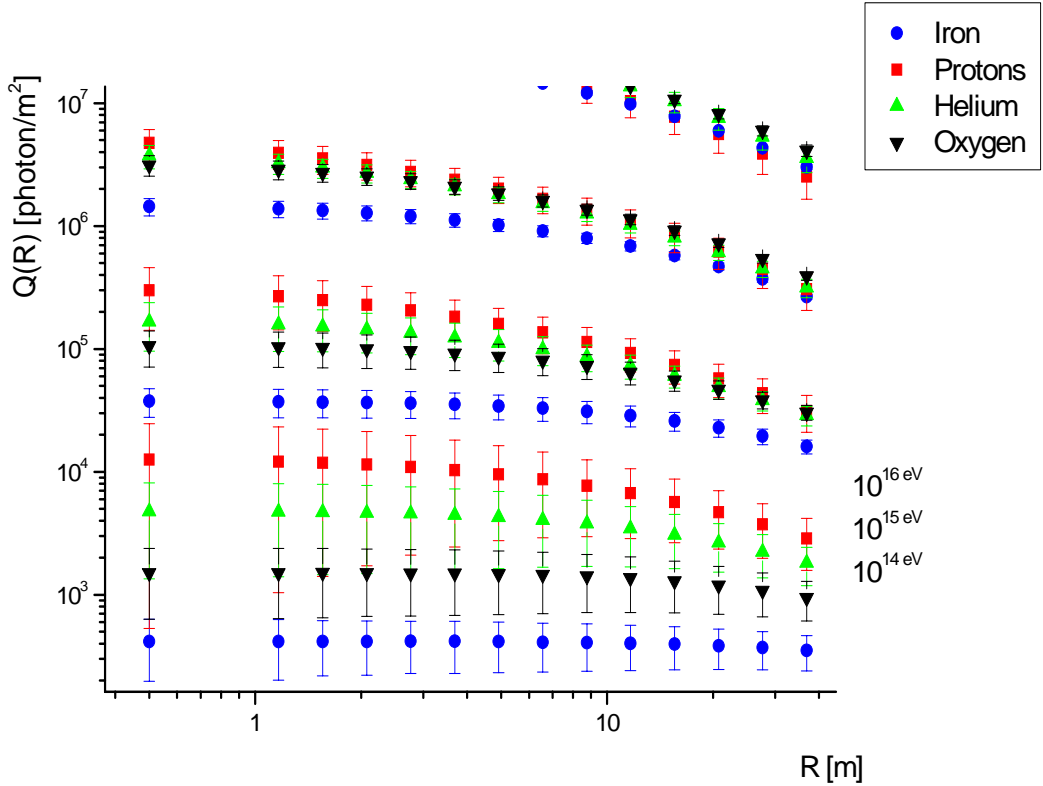

**Fig. 8b** Lateral distributions of Cherenkov light flux in EAS near to the shower produced by primary Protons and Iron, Helium and Oxygen nuclei induced showers in the energy range $10^{15}$ – $10^{17}$ eV at 536g/cm$^2$ observation level

In fig. 8a and 8b are presented the differences between the lateral distribution of Cherenkov light in EAS produced by primary proton, Iron, Helium and Oxygen nuclei. Generally the difference is significant near to the shower axis and in the low energy region i.e. around energies of some $10^{14}$ eV. At the end of the distribution the difference is smaller. However the significant differences observed in the fluctuations are well seen, especially the gradient of the distribution. Obviously in the energy range below the "knee" the obtained lateral distributions of Cherenkov light flux in EAS initiated by primary nuclei is between the Cherenkov light flux generated by proton and iron nuclei as incoming showers.

Similar simulations are carried out for gamma quanta incident particles with practically the same data sets in CORSIKA code [11]. The obtained lateral distributions of Cherenkov light in EAS produced by incoming gamma quanta are presented in fig. 9a in the energy range $10^{10}$-$10^{13}$ eV and fig. 9b $10^{13}$-$10^{16}$ eV. These simulations are very important in attempt to build a useful method for separation of gamma quanta initiated showers from hadron initiated showers. One difference comparing to the lower observation levels [27, 32, 33] is that one can not observe a typical for low observation levels hump i.e. the characteristic ring of Cherenkov photons which appears between 90-120 m from the shower axis in gamma quanta induced showers. This is due essentially to the high mountain observation level and thus the not so important influence of the atmosphere layer to the refractive index.

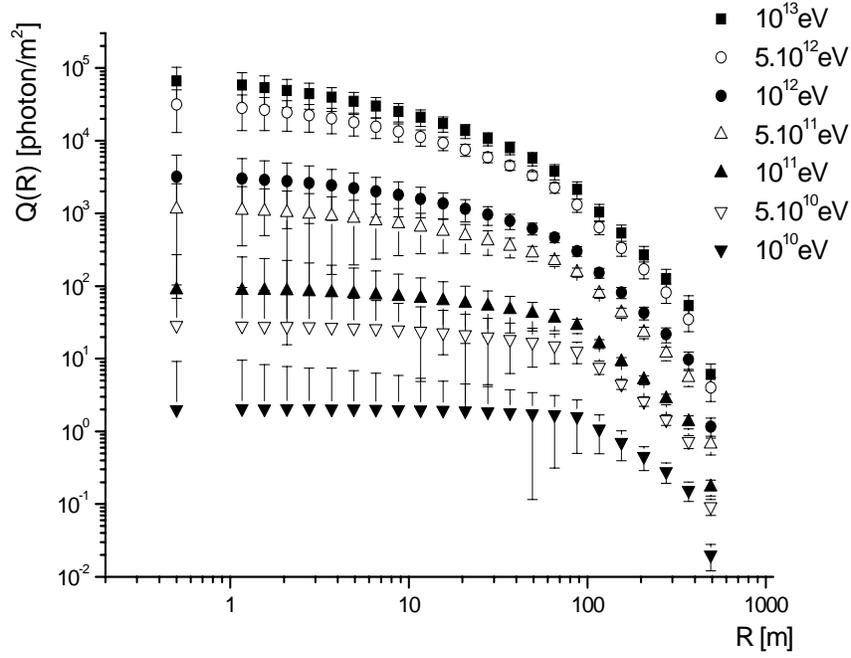

**Fig. 9a** Lateral distributions of Cherenkov light flux in EAS produced by primary gamma quanta induced showers in the energy range $10^{10} - 10^{13}$ eV at 536g/cm$^2$ observation level

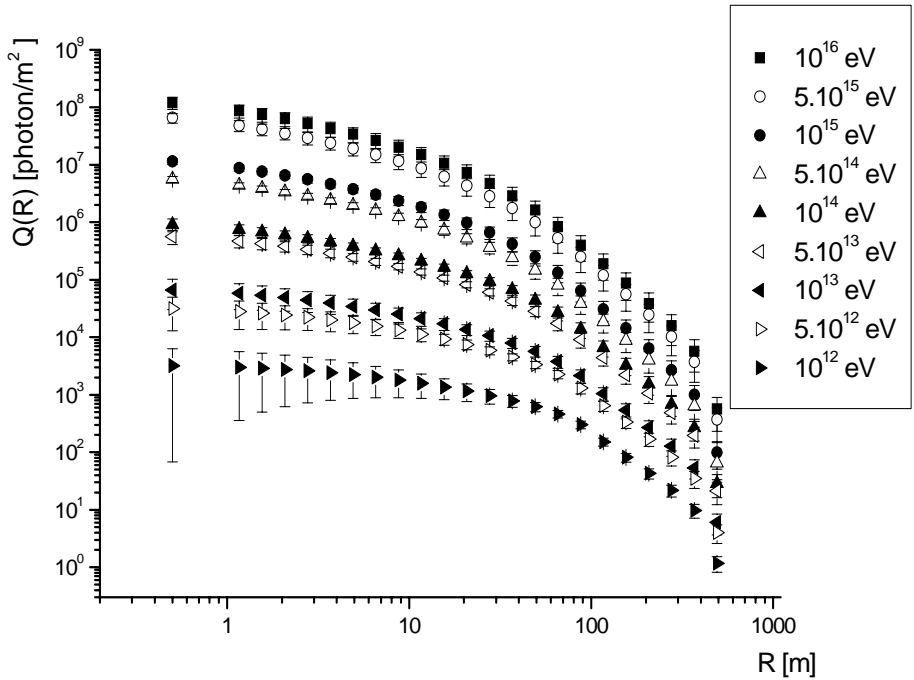

**Fig. 9b** Lateral distributions of Cherenkov light flux in EAS produced by primary gamma quanta induced showers in the energy range $10^{13} - 10^{16}$ eV at 536g/cm$^2$ observation level

Comparing the lateral distribution of Cherenkov light flux in EAS generated by primary protons and gamma quanta one can see that the lateral distribution produced by primary nuclei is wider and with larger density fluctuations as was expected. The difference between lateral distributions of Cherenkov light flux in EAS initiated by proton and gamma quanta showers in the energy range $10^{11}$-$10^{13}$ eV is presented in fig. 10. One can explain this difference counting on the fundamental difference between electromagnetic and hadronic induced showers. In one hand EAS generated by primary gamma quanta is practically pure electromagnetic cascade. Thus the Cherenkov light in EAS initiated by gamma quanta comes essentially from electrons and as a result the lateral distribution of Cherenkov light is much uniform. On the other hand the Cherenkov light in hadron induced showers comes essentially from electromagnetic sub showers initiated by secondary $\pi^0$ decays. The rest comes from charged pions and essentially from the decay muons. One part of the muons can reach the ground i.e. the observation level and thus generate Cherenkov light flux near to the detector. Taking into account that pions generally have large transverse momentum with large fluctuation one can explain the chaotic shower shape and so the larger fluctuations of Cherenkov light flux.

Practically the same simulations are carried out for Carbon, Nitrogen, Calcium and Silicon primary nuclei. The obtained lateral distributions of Cherenkov light in EAS are presented in fig. 11-14.

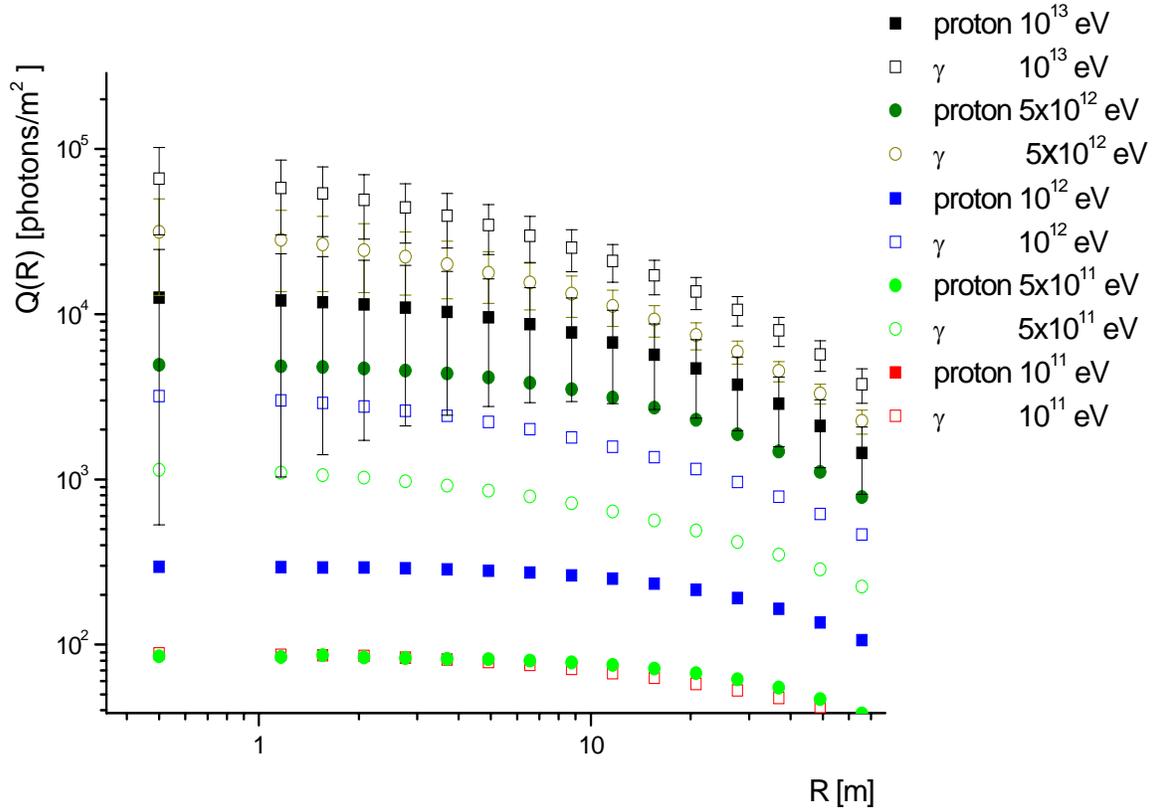

**Fig. 10** Lateral distributions of Cherenkov light flux in EAS produced by primary proton and gamma quanta induced showers in the energy range $10^{11}$ –$10^{13}$ eV at 536g/cm$^2$ observation level

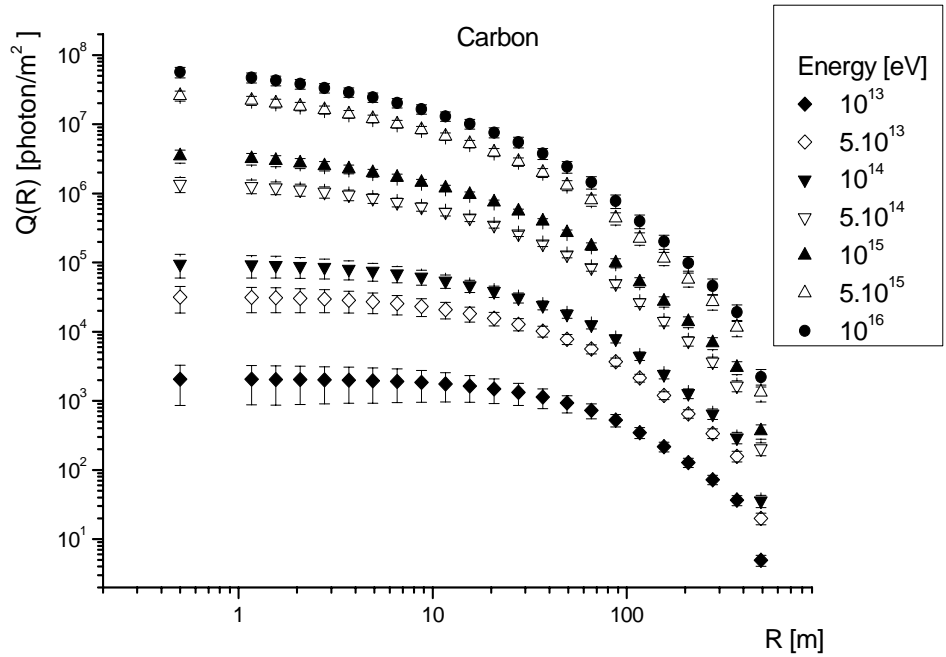

**Fig. 11** Lateral distributions of Cherenkov light flux in EAS produced by primary Carbon nuclei induced showers in the energy range $10^{13}$ –$10^{17}$ eV at 536g/cm$^2$ observation level

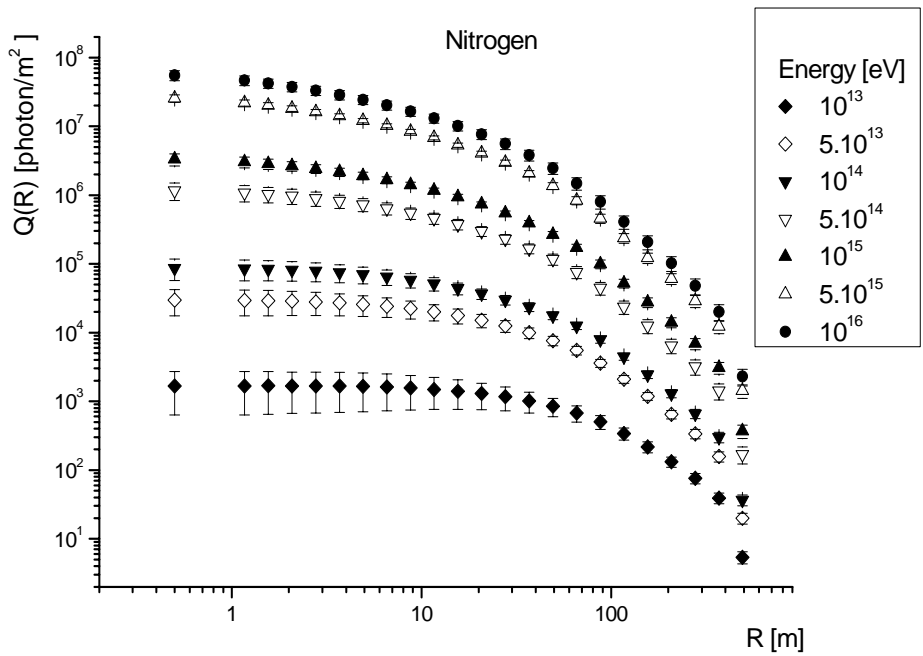

**Fig. 12** Lateral distributions of Cherenkov light flux in EAS produced by primary Nitrogen nuclei induced showers in the energy range $10^{13}$ –$10^{17}$ eV at 536g/cm$^2$ observation level

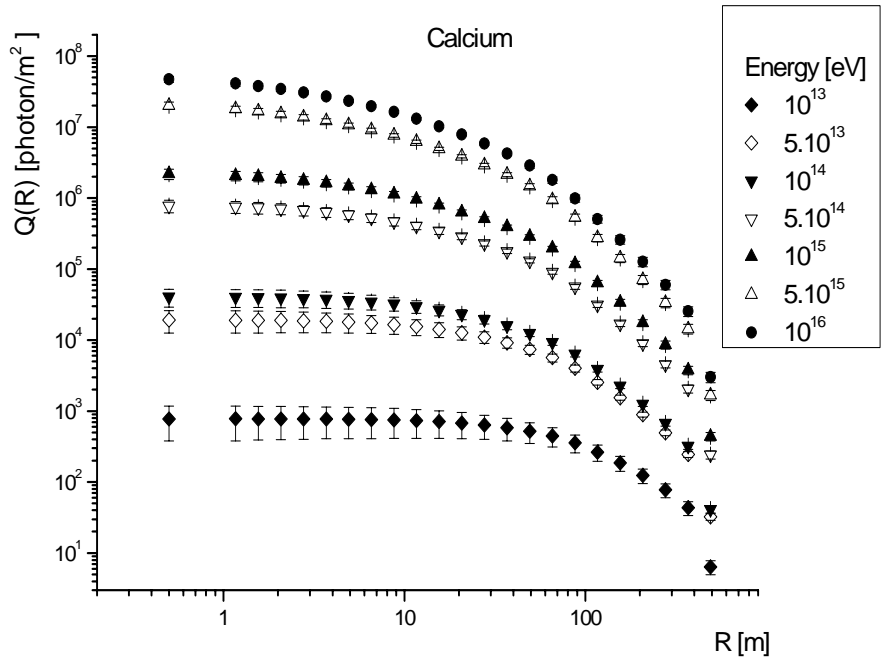

**Fig. 13** Lateral distributions of Cherenkov light flux in EAS produced by primary Calcium nuclei induced showers in the energy range $10^{13}$–$10^{17}$ eV at 536g/cm² observation level

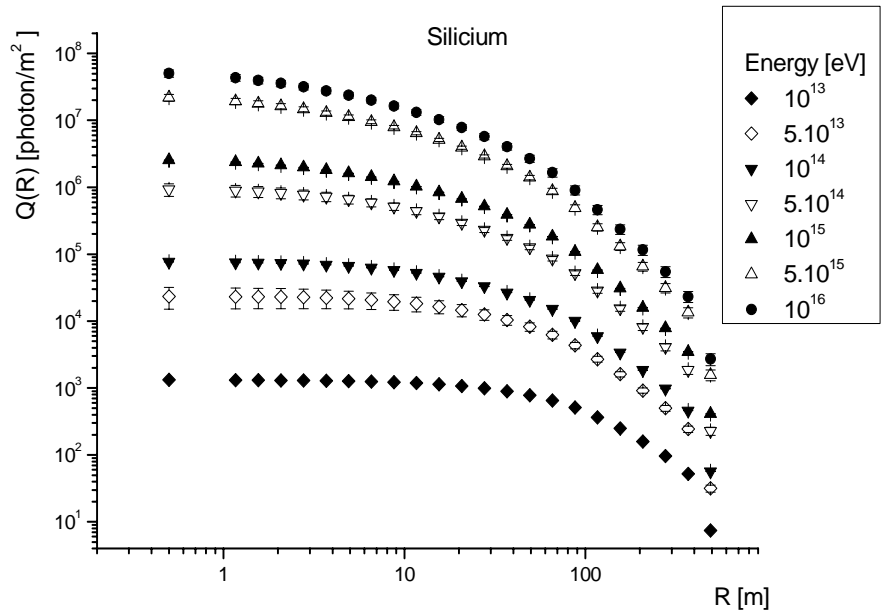

**Fig. 14** Lateral distributions of Cherenkov light flux in EAS produced by primary Silicon nuclei induced showers in the energy range $10^{13}$–$10^{17}$ eV at 536g/cm² observation level

The final aim is to build a databank of Cherenkov light flux lateral distribution in EAS produced by different nuclei and gamma quanta towards to check the previously proposed method for mass composition and energy estimation of primary cosmic ray for large diversity of particles. Moreover the obtained distributions are good basis for rejection of hadronic showers from electromagnetic ones using similar method that proposed by F. Arqueros in [34].

All of the simulated lateral distributions of Cherenkov light in EAS produced by primary hadrons are presented in fig. 15-18. As was expected these results confirm the results presented in fig. 7, 8a and 8b and in [35]. The shape of the distributions is very similar with differences of the density values and the slope. At the same time the relative fluctuations varies as a function of the energy of the initial primary and as well the type.

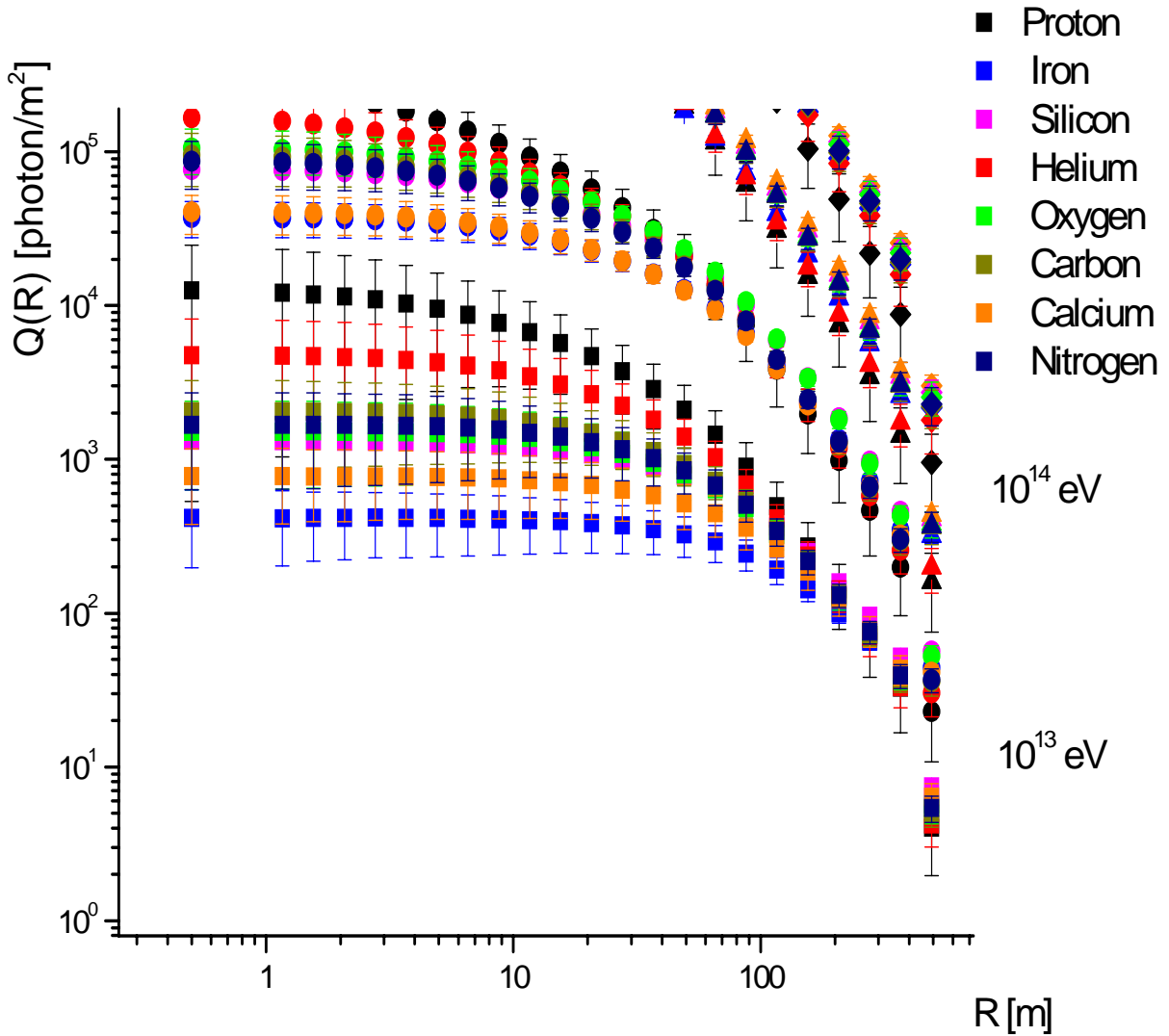

**Fig. 15** Lateral distributions of Cherenkov light flux in EAS produced by primary Proton, Helium, Carbon, Nitrogen, Oxygen, Calcium, Silicon and Iron nuclei induced showers in the energy range $10^{13}$–$10^{14}$ eV at 536g/cm$^2$ observation level

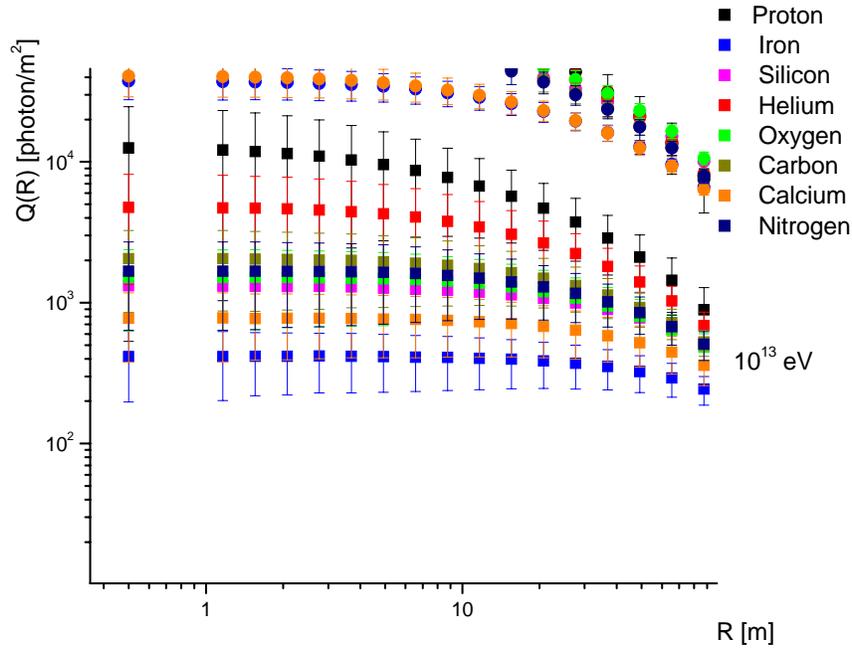

**Fig. 16** Lateral distributions of Cherenkov light flux in EAS produced by primary Proton, Helium, Carbon, Nitrogen, Oxygen, Calcium, Silicon and Iron nuclei induced showers in the energy range $10^{13}$ eV at 536g/cm$^2$ observation level

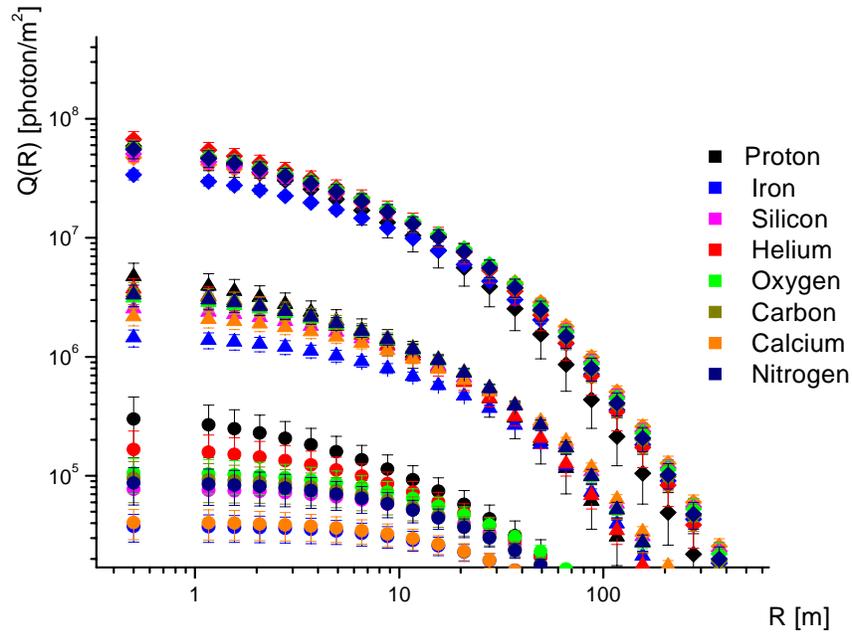

**Fig. 17** Lateral distributions of Cherenkov light flux in EAS produced by primary Proton, Helium, Carbon, Nitrogen, Oxygen, Calcium, Silicon and Iron nuclei induced showers in the energy range $10^{14}$–$10^{16}$ eV at 536g/cm$^2$ observation level

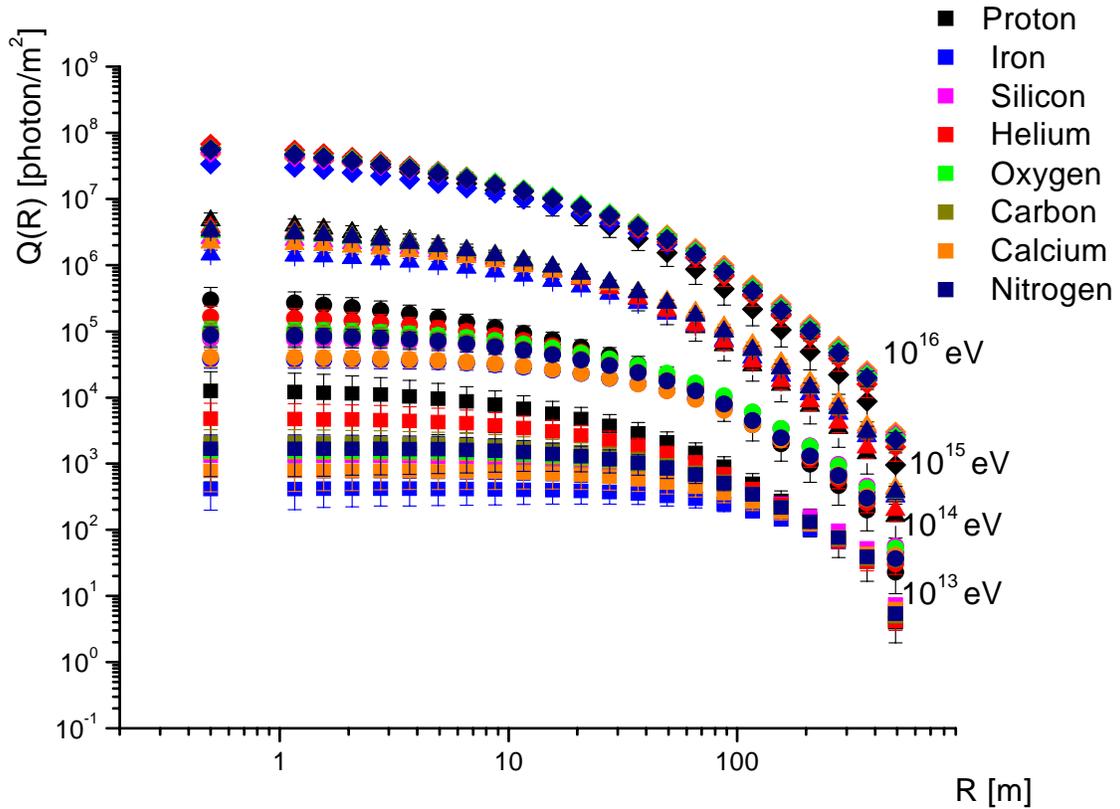

**Fig. 18** Lateral distributions of Cherenkov light flux in EAS produced by primary Proton, Helium, Carbon, Nitrogen, Oxygen, Calcium, Silicon and Iron nuclei induced showers in the energy range $10^{13}$–$10^{16}$ eV at 536g/cm$^2$ observation level

In the near future we will complete the databank including new particles such as Berilium, Borum and Lithium.

## 4. Conclusions

Using CORSIKA 6.003 code [11] and GHEISHA [22] and QJSJET [23] hadronic interaction models the lateral distribution of Cherenkov light in EAS was obtained at Chacaltaya observation level of 536g/cm$^2$. The obtained distributions are obtained in wide energy range between $10^{11}$ and $10^{17}$ eV generated by different primaries, precisely Proton, Helium, Carbon, Nitrogen, Oxygen, Calcium, Silicon and Iron nuclei and primary gamma quanta. The obtained distributions are not experiment dependent and can be useful in practice for future experiment such as HECRE [17] for detector response simulations, checking the reconstruction techniques and methods etc...

The shape of the distributions is discussed and the observed differences as well. The influence of several assumptions during the simulations is also discussed. The results are good basis for solution of the different problems in astroparticle physics such as estimation of the mass composition and energy of the primary cosmic ray and rejection of the hadronic showers from electromagnetic ones which is very important for ground based gamma ray astronomy. In the future the data will be completed within other primaries and observation levels.

## Acknowledgements

We warmly acknowledge our colleagues from IT division in INRNE. We are grateful to the tem of BEO Moussala. This work is partially supported under NATO grant EAP.RIG. 981843.